\documentclass[preprint]{aastex}

\newcommand{\lbol}{L$_{bol}$}
\newcommand{\llyc}{L$_{LyC}$}
\newcommand{\lk}{L$_{K}$}
\newcommand{\nsn}{$\nu_{SN}$}

\newcommand{\CO}{$^{12}$CO}
\newcommand{\htwo}{\mbox{H$_2$}}
\newcommand{\arasa}[2]{Ann. Rev. Astron. Astrophys., #1, #2.}

\newcommand{\apjj}[2]{Ap. J., #1, #2.}
\newcommand{\apjjs}[2]{Ap. J. Supp., #1, #2.}
\newcommand{\apjjl}[2]{Ap. J. (Letters), #1, #2.}

\newcommand{\asa}[2]{Astron. Astrophys., #1, #2.}

\newcommand{\asas}[2]{Astron. Astrophys. Suppl., #1, #2.}
\newcommand{\mn}[2]{M.N.R.A.S., #1, #2.}

\newcommand{\ajj}[2]{A. J., #1, #2.}

\newcommand{\vol}[1]{1}

\newcommand{\solar}{L$_{\odot}$\ }
\newcommand{\solars}{L$_{\odot}$\ }

\newcommand{\solm}{M$_{\odot}$}

\newcommand{\rf}{\par\noindent\hangindent 15pt {}}
\usepackage{graphicx}

\slugcomment{Accepted for publication in the Astrophysical Journal}

\shorttitle{The starburst nucleus of WR galaxy NGC~6764}
\shortauthors{Schinnerer, Eckart \& Boller}

\begin{document}


\title{The Young Starburst Nucleus of the Wolf-Rayet LINER Galaxy NGC~6764}


\author{E. Schinnerer} 
\affil{Astronomy Department, California Institute of Technology, MS
       105-24,Pasadena, CA 91101, USA}
\email{es@astro.caltech.edu}

\author{A. Eckart}
\affil{Universit\"at zu K\"oln, I.Physikalisches
Institut, Z\"ulpicherstra\ss e 77, 50937 K\"oln, Germany}

\and

\author{Th. Boller}
\affil{MPI f\"ur Extraterrestrische Physik, 85740 Garching, Germany}




\begin{abstract}
Near-infrared $K$ band imaging spectroscopy of the central 8'' (1.3~kpc)
in the Wolf-Rayet LINER
galaxy NGC~6764 shows that the most recent star formation is 
most likely
still unresolved at sub-arcsecond resolution ($<$ 100~pc).
The $K$ band continuum source has a size of about 1.5'' (240~pc).
In addition to stellar CO and Na absorption lines as well as
recombination lines of H and He, the $K$ band spectrum shows several
strong emission lines from molecular hydrogen (\htwo).
The H$_2$ line emission is spatially and spectrally resolved showing a
rotating ring/disk of $\sim$ 1.2'' (200~pc) diameter.
\\
An analysis of the nuclear (3'' $\sim$ 480~pc) stellar light using population 
synthesis models in conjunction with NIR spectral synthesis models suggests 
following star formation history:
Two starbursts with decay times of 3~Myr occurred 3 to 5 Myr and 15 to about 
50 Myr ago. Continuous star formation with a SFR of $\sim$ 0.3 \solm yr$^{-1}$ ~
over at least 1~Gyr can also explain the observed parameter.
However, the mass relocation and consumption involved as well as the 
different spatial distribution of the lines associated
with the star formation strongly favor
the 'two starburst' scenario.
In that scenario, up to 35\% of the total observed Br$\gamma$ flux could still 
be due to the AGN,
depending on the assumed age of the older starburst event. 
In contrast to other starburst galaxies (e.g. M82, NGC 7552), the younger 
starburst in NGC~6764 is very likely
located closer to the nucleus and surrounded by the older starburst.
One possible explanation can be that the stellar bar
still transports gas down to radii close to the nucleus. This suggests
that the massive star formation activity is directly competing with the AGN for
the fuel.
\\
We also present the results of a 44 ksec. HRI ROSAT exposure.
The HRI data show the presence of a X-ray source (probably an AGN)
which varies by more than a factor of 2 over a time scale of 7 days.
This implies the presence of a compact source with a discrete or at
most 1000 AU source size. In addition, we find an extended X-ray component
which looks similar to the radio continuum emission
presented in published VLA maps.
Both data sets confirm the composite nature of the center of NGC~6764:
the presence of a compact AGN as well as recent violent nuclear star formation.
\end{abstract}


\section{INTRODUCTION}

In Wolf-Rayet (WR) galaxies the presence of the broad HeII
$\lambda$ 4686 feature (WR feature) signals very recent massive star
formation (Armus, Heckman \&  Miley 1988; Conti 1991).
WR galaxies therefore offer the unique possibility to
study high mass star formation (Maeder \& Conti 1994), since the progenitors of 
WR stars have initial masses $>$ 20 \solm ~and the WR stars themselves are 
younger than 10 Myr. However, it is often
unclear whether the WR line emission is originating from a single isolated HII 
region or is associated with the nucleus of the parent galaxy itself. 
So far, 139 objects are listed in the new Wolf-Rayet (WR) galaxies catalog
(Schaerer, Contini \& Pindao 1999), as 
Wolf-Rayet emission features are rarely observed in galaxies. 
The study of WR galaxies is therefore
essential for understanding the starburst phenomenon in galaxies, as
it offers the unique opportunity to analyze the stellar population at
a well-defined evolutionary stage.
\\
NGC~6764 is a nearby (32~Mpc for $H_0$=75 km s$^{-1}$; 
1$^{\prime\prime}$=160~pc)
S-shaped barred spiral galaxy (SBb), classified as
a LINER galaxy based on optical spectroscopy (Osterbrock \& Cohen 1982).
NGC~6764 is unusual because it also displays
a prominent 466 nm Wolf-Rayet emission feature
at the nucleus (Osterbrock and Cohen 1982) and was already in the
first WR catalog of Conti (1991).
The galaxy contains a nuclear stellar optical continuum 
with a ''width'' of $\sim 1.6"$ or about 260~pc
(Rubin, Thonnard, \& Ford 1975).
\\
Eckart et al. (1991) presented single dish measurements of the
J=1-0 and J=2-1 rotational transitions of $^{12}$CO and $^{13}$CO obtained 
with the IRAM 30m telescope, as well as the 
first $JHK$ images of NGC~6764. Eckart et al. (1996) presented
X-ray observations, high resolution optical, millimeter-wave
interferometer measurements of the 115 GHz $^{12}$CO(1-0) line emission,
the first imaging observations of the 2.12 $\mu$m H$_2$ emission, and
measurements of the 2.06 $\mu$m He~I and the 2.17 $\mu$m Br$\gamma$ line emission
toward the nucleus of NGC~6764.
Their  NIR and optical spectroscopic data which include the
466~nm Wolf-Rayet feature
revealed a dense concentration of molecular gas in the center
and a very recent (few 10$^7$ yr) starburst at the nucleus of NGC~6764.
This starburst is most likely responsible for
the presence of a few thousand Wolf-Rayet stars in the nucleus of
NGC~6764 (Eckart et al. 1996).
\\
\\
In this paper we present the first $K$ band integral field spectroscopy
and ROSAT X-ray observations using the high-resolution imager
(HRI) of the nucleus of NGC~6764. We describe our observations in \S 2,
and present an analysis of the X-ray data in \S 3.
In \S 4 we discuss the result from the integral field spectroscopy,
and in \S 5 we analyze the nuclear star formation history of NGC~6764. 
We present a brief summary in \S 6.

\section{OBSERVATIONS}

\subsection{$K$ band integral field spectroscopy}

NGC~6764 was observed in the $K$ band (2.2 $\mu$m) on 1997, July 26 and 28
with the MPE integral field spectrograph 3D (Weitzel et al. 1996) combined
with the tip-tilt corrector ROGUE (Thatte et al. 1995) at the 3.5~m telescope
in Calar Alto, Spain. The image scale was 0.5''/pixel. The total
integration time on source was 2880~s. Two fields off-set by 1.5'' in east-west
direction were combined to cover the nuclear region.
Since the seeing was different in both nights
($\le$ 1'' on July 26 and $\sim$ 1.2'' on July 28), the data set from
July 26 was convolved to the resolution of the second night for the
combined data cube with a larger field of view. 
\\
3D obtains simultaneous spectra for each pixel
(R$\equiv\lambda/\triangle\lambda\approx$
750 for $K$ band) of an 8'' $\times$ 8'' field. This is done using
an image slicer which rearranges the two-dimensional focal plane onto a 
long slit of a grism. The dispersed spectra were then detected on a NICMOS~3 
array. A detailed description of the instrument and the data reduction is
given in Weitzel et al. (1996). The data reduction procedure converts each 
two dimensional image into a three dimensional data cube with two spatial 
and one spectral axes. The data cubes were co-added and centered on
the continuum peak. All images were dark-current and sky-background
subtracted, corrected for dead and hot pixels, and spatially and
spectrally flat-fielded. To correct for the effects of the Earth's
atmosphere on the $K$ band spectrum, a standard star was observed. This
standard spectrum was first divided by a template spectrum of the same
spectral type (Kleinmann \& Hall 1986) in order to remove stellar
features. The effect due to different zenith distance of the source and
the standard star was minimized using the ATRAN atmospheric model (Lord
1992), mainly to correct for the different atmospheric absorption. The
source data were then divided by the atmospheric transmission spectrum.
\\
We flux calibrated the data adopting the integrated $K$ band flux density value
given by Rieke (1978) in a 8.5'' aperture centered on the nucleus. 
Due to the different spatial resolution and S/N each data cube was flux
calibrated individually before combination.
A comparison to the line fluxes of Eckart et al. (1996) in a 5''
aperture showed good agreement within the errors 
for the HeI~$\lambda$2.06$\mu$m and
H$_2$S(1)~$\lambda$2.12$\mu$m lines. For Br$\gamma ~\lambda$2.16$\mu$m
the obtained flux was by a factor of 2 higher. This is probably related to the
fact that the Br$\gamma$ line has a weak broad component (FWZI $\sim$
2100 km/s).

\subsection{ROSAT HRI observations}

NGC 6764 was observed using the HRI detector (David et al. 1996)
on board the ROSAT satellite (Tr\"umper 1983) between
1995 September 2 (10:23:45 UT) and 1995 September 19 (10:23:38 UT).
The total exposure time was 44660 s and the source was centered
on-axis in the field of view. 
The centroid position of NGC 6764 in the HRI image, computed
from a Gaussian fit to the spatial distribution is
$\alpha_{2000} = 19^h08^m16.^s22 \pm 0.03s$,
$\delta_{2000} = +50^o56^{'}2.20^{''} \pm 0.59^{''}$.
The internal ROSAT HRI pointing error is of the order of $\pm$ 5 ''.
The source cell size was obtained from a radial profile of counts 
within a ring centered on the centroid position of NGC 6764.
For distances larger than 25'' from the centroid position the 
radial profile is consistent with the background level.
The number of source plus background photons within a radius of 25''
is 457 $\pm$ 21. The background was determined in a ring around the
source with an inner radius of 40'' and an
outer radius of 80''. 
The number of background photons normalized to the source
cell size is 109 $\pm$ 10. The resulting number of source photons
is therefore 348 $\pm$ 25 corresponding to a mean count rate of 
$\rm (7.8 \pm 0.3) \times 10^{-3} counts\ s^{-1}$.

\section{ROSAT HRI RESULTS}

Our HRI data show both the presence of an X-ray source that strongly varies on
the time scale of several days and an additional extended X-ray component
that looks similar to published VLA radio continuum maps.
Both findings confirm the composite nature of the center of NGC~6764:
The presence of a compact active nucleus and recent nuclear star formation
(see also Gon\c{c}alves, Veron-Cetty \& Veron 1999).

\subsection{X-ray Variability}

\begin{figure}[t!]
\begin{center}
\includegraphics[height=15.5cm,width=10.5cm,angle=-90.]{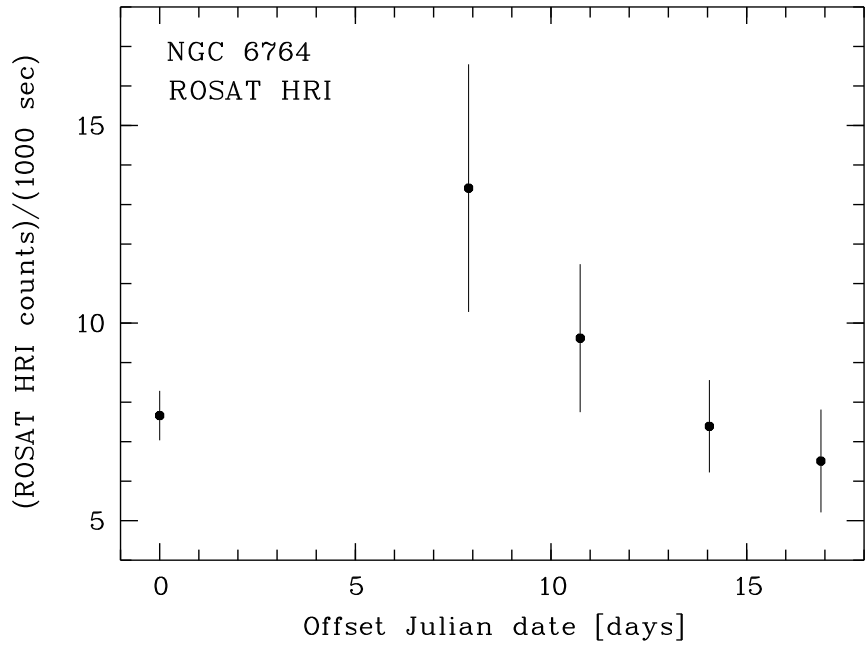}
\end{center}
\figcaption{\label{xxa01}  
ROSAT HRI light curve of NGC 6764. The abscissa label gives 
the offset Julian date. Each data point is plotted in the middle of the
exposure interval from which it was obtained, and the sizes of the
exposure intervals lie within the data points themselves. NGC 6764 shows
indications for count rate variations by a factor of about 2 within
a few days (see text for details).
}
\end{figure}

For the extragalactic sources studied by Walter \& Fink (1993) with the 
ROSAT PSPC the weighted mean photon index is 2.3 with an uncertainty of
0.03. Due to the limited photon statistics the PSPC observations of
the ROSAT All-Sky Survey of NGC~6764 do not constrain the spectral
X-ray properties. Therefore,
we have used the spectral shape of a simple power-law model
with an photon index $\rm \Gamma$ of 2.3 and an absorbing column
equal the Galactic absorption by neutral hydrogen of
$\rm 6.0 \times 10^{20}\ cm^{-2}$ towards NGC 6764 (Dickey \& Lockman 1990).
We derive that the mean HRI count rate
of $\rm 7.8 \times 10^{-3} counts\ s^{-1}$
corresponds to an absorption-corrected 0.1-2.4 keV flux of 
$\rm 7.13 \times 10^{-13} erg\ cm^{-2}\ s^{-1}$ (see Fig. \ref{xxa01}).
Using equation (7) of Schmidt \& Green (1986) (with $\rm H_0 = 75 km\
s^{-1}\ Mpc^{-1}$ and $\rm q_0 = \frac{1}{2}$) we find that the mean
HRI count rate of NGC 6764 corresponds to an isotropic luminosity of
$\rm 8.71 \times 10^{40} erg\ s^{-1}$ or about 2.3$\times$10$^{7}$\solars
fully consistent with the luminosity derived by Eckart et al. (1996).
Here we assume that the radiation is isotropically radiated and that
relativistic beaming effects are not relevant for the derived value of
the luminosity. 
Kriss, Canizares \& Ricker (1980) give an upper limit of the X-ray flux of
NGC~6764 of L$_X$$<$ 8.6$\times10^{40}$ergs~s$^{-1}$ over the 
($0.5-4.5$~keV) energy band. 
\\
\\ 
An examination of the arrival times of the X-ray photons shows 
deviations from the mean count rate and suggests the presence
of variability on very short time scales. 
Fig. \ref{xxa01} displays the ROSAT HRI light curve obtained for
NGC 6764 between 1995 September 2 and 19. The exposure times 
for the individual data points range between 1575 seconds 
(Offset Julian date 7.9) and 26520  seconds (Offset Julian date 0.0). 
The integration time per bin is always larger than 1550 seconds to avoid
apparent count rate variations due to the ROSAT wobble.
NGC 6764 shows count rate variations by a factor of about 2
on a time scale of 7 days. 
A constant light curve model can be rejected
with $\rm 97.6$ per cent confidence based on a $\chi^2$ statistic.
The continuous decrease
of the ROSAT HRI count rate starting at day 7.9 of the observations
further accounts for true changes of the X-ray emission in the
distant Wolf-Rayet LINER galaxy NGC 6764.
The presence of an extended X-ray component (discussed in
section \ref{soft}) that accounts for about half of the total
luminosity suggests that the observed variabilitiy (if entirely attributed to
a nuclear source) was significantly higher than a factor of 2.
\\
\\
The minimum count rate of $\rm 5.95 \times 10^{-3} counts\ s^{-1}$ 
was observed at day 1.5 within a 4760 second exposure time interval.
The maximum count rate is $\rm 1.34 \times 10^{-2} counts\ s^{-1}$
observed at day 7.9 within a 1575 s exposure interval. 
The light curve of NGC 6764 shows a continuous decline of the count rate
starting at day 7.9 up to the end of the observations.
Such count rate variations suggest that a major fraction of the 
variable X-ray emission originates close to a super-massive black hole.
Assuming that the variability is not affected by beaming or relativistic
motions the upper limit for the size of the variable X-ray emission
is $\rm R \approx \Delta t\ c \approx 1.6 \times 10^{16} cm$ or only about
10$^3$AU.

\subsection{\label{soft}Extended X-ray emission}

\begin{figure}[t!]
\begin{center}
\includegraphics[height=5.5cm,width=5.5cm,angle=0.]{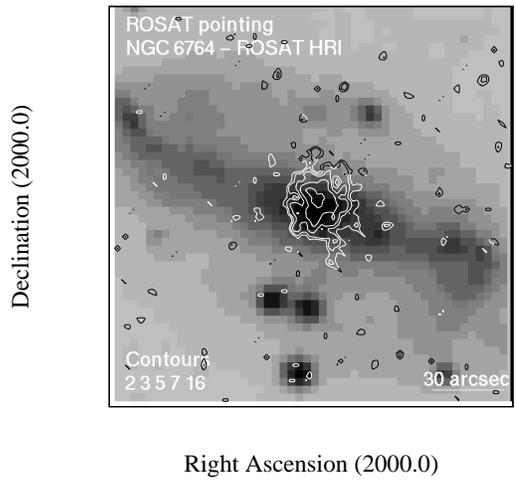}
\end{center}
\caption{\label{xxa02}
Pointed ROSAT HRI image of NGC~6764 in contours
overlaid on an optical  Digitized Sky Survey grayscale image.
The X-ray emission is clearly extended in north-south direction.
}
\end{figure}

The X-ray emission originating from NGC~6764 shows a complex structure:
In addition to a source component (see Fig. \ref{xxa02})
at the nuclear position, we find an extended structure with indications
for different centers of enhanced X-ray emission north-west and south-west
of the central position. The extended emission accounts for about half of the
total X-ray luminosity in the 0.1-2.4 keV band derived from the mean
count rate. If parts of the extended emission are due to a soft X-ray spectrum, 
this suggests that the X-ray luminosity of this extended component 
is either due to Bremsstrahlung emission or an emission line dominated
spectrum. With the ROSAT HRI and also the PSPC observations we are not
able to differentiate between Bremsstrahlung models and an emission
line dominated spectrum from hot diffuse gas based on the model
calculations of Raymond \& Smith (1977). The derived plasma temperatures of 
both models are constistent within the errors. This would be in agreement with 
emission from a very diffuse component in addition to a contribution from
compact sources indicated by our data and  expected from 
supernovae, hot stars, and X-ray binaries.
Eckart et al. (1996) have shown that the X-ray luminosity could be
entirely explained by the energy input of supernovae, and that the total
mass of hot gas probably amounts to several 10$^6$\solm.
\\
Starburst galaxies show typical values of 10$^{-2}$ to 10$^{-4}$ for
their L$_X$/L$_{FIR}$ ratios (Boller 1999). The L$_X$/L$_{FIR}$ ratio for 
NGC~6764 lies within the range expected for starburst, if we use
the X-ray luminosity for the extended component of L$_X$ $\sim$ 1.15 
$\times$10$^7$ \solar ~and a total far-infrared luminosity of 
L$_{FIR}$ $\sim$ 2$\times$10$^{10}$ \solar. As discussed in section
\ref{xxlbol} about 40\% of the far-infrared luminosity can be attributed
to the nuclear starburst (inner $\sim$ 250~pc). This leaves still 
enough luminosity for the extended
component. Therefore, we find that the L$_X$/L$_{FIR}$ ratio of the
extended X-ray emission in NGC~6764 is in agreement with those found
for starburst galaxies. The spatial extent of the extented emission of
$\sim$ 10'' FWHM ($\sim$ 1.6 kpc) is similar to the extent of
starburst rings of about 1 to 2 kpc.
\\
We also note that the extended X-ray structure is 
similar to the structure found in the VLA radio maps 
(Ulvestad, Wilson, \& Sramek 1981; Wilson \& Willis 1980).
Baum et al. (1993) discussed the origin of the radio emission
suggesting strong star formation and due to the alignment of the radio
emission with the minor kinematic axis pressure driven outflow.
A large-scale H$\alpha$ map shows extended nuclear H$\alpha$ emission
over the inner 10'' (D. Frayer, private communication).
This suggests that the hot X-ray emitting gas and the synchrotron radiation
emitting relativistic gas component are spatially coexisting and probably
both are a result of strong recent circum-nuclear star formation.

\section{$K$ BAND INTEGRAL FIELD SPECTROSCOPY RESULTS}

\begin{figure}[t!]
\begin{center}
\includegraphics[height=15.5cm,width=10.5cm,angle=-90.]{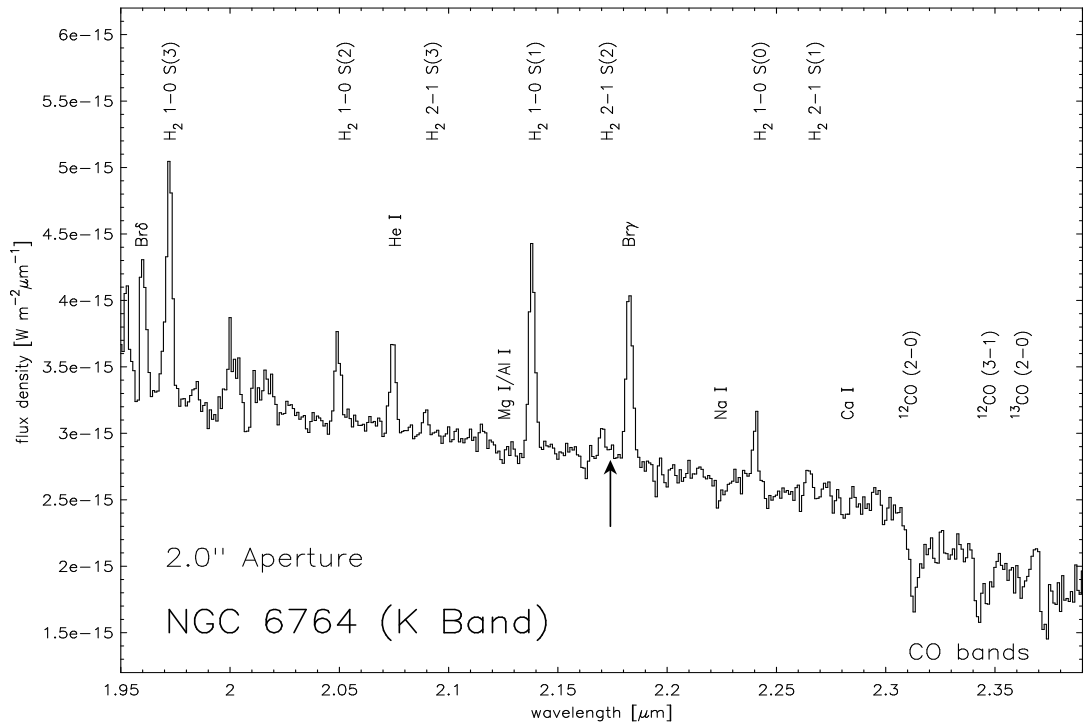}
\end{center}
\figcaption{\label{xxa03}
$K$ band spectrum of the inner 3'' of NGC~6764. The
detected lines are indicated. The arrow marks the position of the
HeI/HeII line complex discussed in section \ref{xxwr} (see also Fig.
\ref{xxa06}).
}
\end{figure}

The nuclear $K$ band spectrum is dominated by emission lines from warm
molecular \htwo. The slope of the continuum indicates no strong
dilution from non-stellar sources like dust emission or AGN power law
contribution. 
The $K$ band continuum map gives a source size of about FWHM 2.2''
$\times$ 1.7'' (PA $\sim$ 95$^o$), larger than the seeing of 1.2''. 
This implies that the $K$ band continuum source is extended with 
an intrinsic source size of about 1.7'' $\times$ 1.2'' (PA $\sim$ 95$^o$).
This is quite consistent with the July 26 data, where we obtained an uncorrected
source FWHM of 1.8'' $\times$ 1.3'' (PA $\sim$ 95$^o$). 
\\
We made line maps for all the lines indicated in Fig. \ref{xxa03} by
averaging the adjacent continuum channels and subtracting the resulting mean
from all channels.
These line channels were subsequently co-added. All fluxes and
equivalent widths (EWs) are measured in circular
apertures, the corresponding errors are the standard deviations (1$\sigma$)
for the given apertures in the line maps (Tab. \ref{xxt01} and
\ref{xxt03}) and do not include calibration uncertainties 
or systematic effects which we estimate to be of the order of 10\%.
All calculations for the stellar population synthesis are made for a 3'' 
(480~pc) aperture.

\subsection{\label{xxheh}The H and He Recombination Lines}

\begin{figure}[t!]
\begin{center}
\includegraphics[height=12.5cm,width=12.5cm,angle=0.]{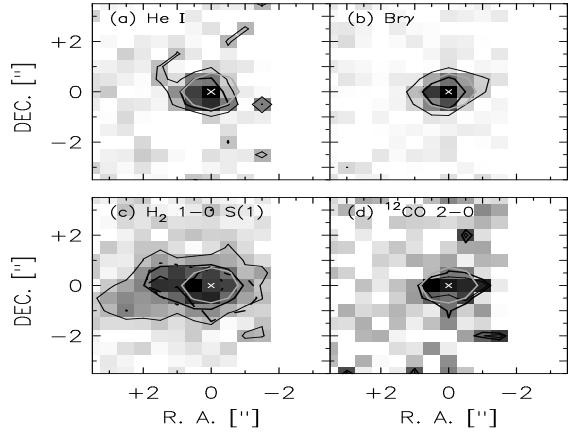}
\end{center}
\figcaption{\label{xxa05}
Maps of the (a) HeI, (b) Br$\gamma$ and (c) \htwo 1-0 S(1) line emission as well
as the (d) $^{12}$CO (2-0) absorption line from the July 26 data cube. The
50\% contours of the lines (fat black line) and the adjacent continuum
(fat grey line) are indicated as well as the 2$\sigma$ contour (thin
black line). The blue (dashed contour) and red (dotted contour) wing
of the \htwo 1-0 S(1) line emission are given as well.
}
\end{figure}

The Br$\delta$, Br$\gamma$ and HeI
line emission is spatially unresolved at our resolution. The comparison of
the FHWM in the line maps and the adjacent continuum channels shows
clearly that the extent of the emission lines is more compact than the
continuum emission (Fig. \ref{xxa05}). This indicates 
that the line emission is arising in a region considerably smaller than 150~pc
(about 1'').
The ratio of Br$\delta$/Br$\gamma$ $\sim$ (0.61$\pm$0.12) is
consistent with values of 0.65-0.67 expected for 'case B recombination' 
(Osterbrock 1989) which suggests that extinction only plays a minor role at these wavelengths (see also Eckart et al. 1996).
\\
Assuming that the HeI 2.06 $\mu$m and Br$\gamma$ line emission is
arising in HII regions around hot, young stars,
the HeI/Br$\gamma$ ratio can be used to estimate the mean effective 
temperature T$_{eff}$ of these ionizing stars. We observe a ratio of 0.46
$\pm$ 0.04 indicating T$_{eff}$ $\sim$ 35 000 K, representative of an O8
star dominated ionizing stellar population. Note that we neglected effects 
like mixed-in dust and differences in the velocity structure of the Str\"omgren
spheres (see Lan\c{c}on \& Rocca-Volmerage 1996, Doherty
et al. 1995). Since NGC~6764 is classified as a LINER, we could expect some
AGN contribution to the Br$\gamma$ line emission, making the effective
temperature a lower limit (see also section \ref{xxsb}). 
This is consistent with the finding 
that about 66\% (72\%) of the nuclear
H$\alpha$ (H$\beta$) line emission which is coming from a narrow line
component is of stellar origin (i.e. consistent with HII region line ratios;
Gon\c{c}alves, Veron-Cetty \& Veron 1999).

\subsection{The \htwo ~Emission Lines}

The \htwo 1-0 S(1) line is the strongest of all the six \htwo
~lines detected with a S/N $\geq$ 3 in the $K$ band spectrum. The 
\htwo 1-0 S(1) emission
is extended in the east-west direction with a FWHM of $\sim$ 3.3''
(500~pc corrected for seeing) in the line maps (see also 
Eckart et al. 1996). The extent is even larger than that in the $K$ band 
continuum of 2.2''. The
maximum of the line emission is off-set by about 0.25'' east from the
continuum peak. The \htwo 1-0 S(1) line has a FWZI of $\sim$ 950 km/s.
Line maps of the blue and red line wing (centered $\pm$ 270 km/s of
the line center with widths of 400 km/s) show a shift in the emission 
distribution from the south-west to the north-east indicating a rotating 
gas disk or ring. The emission peaks of the blue and red line wings
are about 1.2'' (200~pc) apart. These observed kinematics and distribution
of the warm molecular gas are similar to those of the cold molecular gas 
observed in its \CO ~line emission (Schinnerer et al. in prep.).

\subsection{\label{xxco}The Stellar Absorption Lines}

The CO absorption band heads at $\lambda$2.29 $\mu$m are clearly
visible in the spectrum (Fig. \ref{xxa03}). In addition the NaI
absorption at $\lambda$2.21 $\mu$m is most likely detected, all those lines 
indicate the presence of cool, evolved stars. The spatial extent of the
stellar absorption lines coincides with the extent of the $K$ band
continuum (Fig. \ref{xxa05}), in contrast to the H and He recombination lines 
which are spatially unresolved at our resolution. The situation therefore
indicates that the hot, young stars and the cool, evolved stars are distributed differently.
\\
\\
Using the measured equivalent widths (EWs) of the detected stellar
absorption features and their ratios, it is possible to classify the
mean stellar type (e.g., Origlia, Moorwood \& Oliva 1993, Oliva et al. 1995, 
F\"orster-Schreiber 2000). The $^{12}$CO 2-0/NaI ratio of (4.57$\pm$1.56) is
consistent with effective temperatures T$_{eff}$ between 3000 K and 4800 K. 
The $^{12}$CO(2-0)/$^{13}$CO(2-0) ratio indicates a T$_{eff}$ $\leq$ 4500 K,
since these lines are vanishing for effective temperatures $>$ 4500~K 
(F\"orster-Schreiber 2000).
Assuming that the total $K$ band continuum is due to cool stars and that the
measured equivalent widths (Table \ref{xxt03}) are not diluted, a 
comparison to EW of standard stars (e.g. Origlia, Moorwood, Oliva 1993, 
Schreiber 1998) suggests the presence of either late K supergiants or early M giants.
Under the assumption that there is a
continuum flux density contribution from younger
stars (indicated by the HeI and Brackett emission) and from the AGN, the 
mean stellar type has to be even cooler since in this case
the CO band heads appear diluted and are therefore deeper than 
actually measured. 
The cool, evolved stars can be either
members of a few Myr old starburst (being mostly red supergiants, RSGs) or
part of an older stellar population (being mostly red giants, RGs).

\section{THE NUCLEAR STAR FORMATION HISTORY}

To investigate the nuclear star formation history further, we use the 
population synthesis code STARS (Sternberg \& Kovo, private
communication) which was
successfully applied to a number of galaxies (NGC~1808: Krabbe,
Sternberg \& Genzel 1994, NGC~7469: Genzel et al. 1995, NGC~7552:
Schinnerer et al. 1997, IC~342: B\"oker et al. 1997, Circinus:
Maiolino et al. 1998, M~82: Schreiber 1998, NGC~3227: Schinnerer et al.
2000)
in conjunction with the NIR spectral synthesis code SPECSYN (Schinnerer et al.
1997, 2000). 
\\
The model is similar to other stellar population synthesis
models (e.g. Lan\c{c}on \& Rocca-Volmerange 1996,
Leitherer et al. 1999) and includes the most recent stellar
evolutionary tracks (Schaerer et al. 1993, Meynet et al. 1994). 
A summary of the systematic uncertainties in the population
synthesis models is given in Leitherer et al. (1999); for a discussion of the
AGB phase which is implemented in STARS see Oliva \& Origlia (2000). STARS gives
output observable parameters such as the bolometric luminosity
L$_{bol}$, the $K$ band luminosity L$_K$,the Lyman continuum luminosity 
L$_{LyC}$, and the supernova rate $\nu_{SN}$, as well as the diagnostic
ratios between these quantities:
L$_{bol}$/L$_{LyC}$, L$_K$/L$_{LyC}$ and 10$^{9}\nu_{SN}$/L$_{LyC}$.
All three ratios are measures of the time evolution and the shape of
the IMF, with slightly different dependencies on the exponential slope
$\alpha$ and the upper mass cut-off m$_u$.
Hertzsprung-Russell diagrams (HRDs) representing the distribution of these luminosities are 
calculated. SPECSYN uses the distribution of the $K$ band luminosity L$_K$
within a HRD (from STARS) to weight standard star spectra of 
different spectral type and luminosity (Schinnerer et al. 1997, 2000).
In addition, effects like extinction or non-stellar contributions from
dust emission or an AGN can be included as well.

\subsection{\label{xxlbol}The Parameters for the Population Synthesis}

To estimate {\it \lbol} for the central 3'' ($\sim$ 500~pc) of NGC~6764 
we followed the approach of Eckart et al. (1996). For starburst regions
the FIR luminosity can be assumed to be equivalent to the bolometric
luminosity (Telesco, Dressel \& Wolstencroft 1993). 
Under the assumption that the relative
contribution from the star forming regions is similar at all the IRAS
wavelengths, we scaled the IRAS fluxes by 0.4 at 
12 $\mu$m (0.36 Jy), 25 $\mu$m (1.29 Jy), 60 $\mu$m (6.33 Jy) and 
100 $\mu$m (11.56 Jy) accordingly to the measured 10.6
$\mu$m flux in a 5'' aperture (Rieke 1978). 
Since the 10.6 $\mu$m measurement of Rieke (1978) and the IRAS 12
$\mu$m measurement have similar filter width of $\sim$ 5 $\mu$m and the strong absorption
and emission features in the mid-IR spectrum of NGC~6764 (Rigopoulou
et al. 1999) are covered in both measurements, this approach seems
justified to obtain the nuclear mid- and far-IR emission.
To derive {\it \lbol} the
relation given by Sanders \& Mirabel (1996) was used.
This approach was examined in the case of NGC~7552, there similar values for
\lbol ~were obtained by using different methods (Schinnerer et al. 1997).
\\
{\it \lk} and {\it \llyc} were estimated via the equations given in
Genzel et al. (1995) (using a $\delta \lambda$ = 0.6$\mu$m as the $K$
band width) and taking the values measured with 3D in a 3'' aperture. 
For {\it \nsn} we used the empirical relation between the 5 GHz flux and {\it \nsn}
given by Condon (1992). To derive the actual value, we only used the 
nuclear 5 GHz flux of 3.4 mJy from Baum et al.
(1993), since the H/He recombination lines as well as the stellar
absorption lines are smaller in extent than the nuclear radio
source size. The derived quantities are listed in Tab. \ref{xxt04}.
(All equation are described in Schinnerer, Eckart, \& Tacconi 1998.)

\subsection{\label{xxwr}Evidence from Wolf-Rayet features}

\begin{figure}[t!]
\begin{center}
\includegraphics[height=15.5cm,width=10.5cm,angle=-90.]{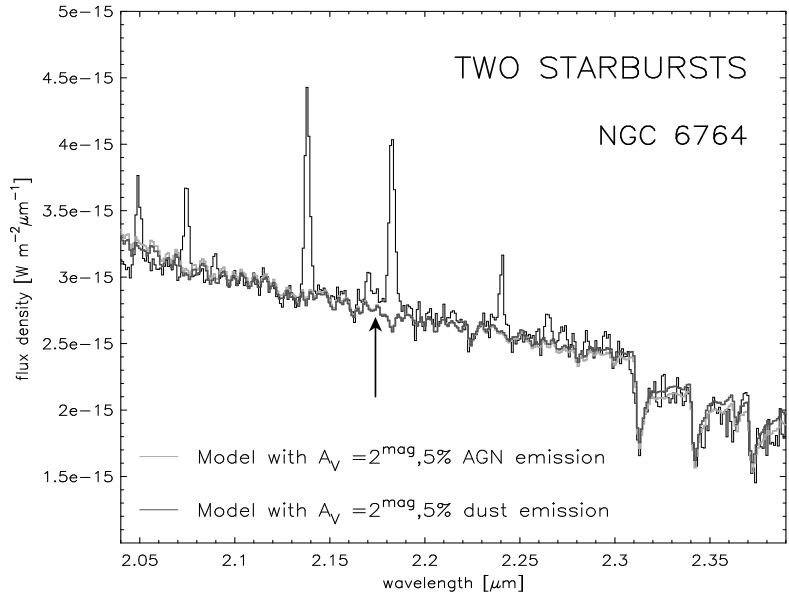}
\end{center}
\figcaption{\label{xxa06}
Comparison of the observed spectrum to the synthesized spectrum for
the case of the two starbursts. We used ages for the two starburst
events of 5~Myr and 25~Myr (see also Table \ref{xxt04}).
The arrow marks the position of the
HeI/HeII line complex discussed in section \ref{xxwr}.
}
\end{figure}

Eckart et al. (1996) found that the WR feature at 
4660 \AA ~(NIII $\lambda$4640 and HeII $\lambda$4686) is spatially
extended. However, the 3D HeI and Br$\gamma$ line maps show no indication
for extended emission which may be due to our sensitivity.
The optical line emission is mainly due to N-rich WR stars (subtype WN)
(Eckart et al. 1996). 
In their recent optical data Kunth \& Conti (1999) were able to identify
a WR feature at CIII $\lambda$5696 and CIV $\lambda$5808 arising from C-rich
WR stars (subtype WC). 
\\
In the $K$ band, the \ion{C}{4} and \ion{C}{3} lines at 
2.08 $\mu$m and 2.11 $\mu$m, respectively, are expected from WC stars 
(Figer, McLean, \& Najarro 1997) which are not detected in the 3D data
at our sensitivity. 
We see indication for the HeI/HeII line complex at 2.16 $\mu$m (Fig.
\ref{xxa03} and \ref{xxa06}) indicative of the presence of WN stars
(Figer, McLean, \& Najarro 1997).
Since in late WN stars this line complex is almost as prominent as the 
2.06 $\mu$m HeI line (Figer, McLean, \& Najarro 1997), this implies
that most of the line emission seen at 2.06 $\mu$m is due to HII
regions and not WR stars.
\\
Population synthesis results of WR populations show that considerable
numbers of WN stars (in addition to WC stars) are present after about 3~Myrs
(Schaerer \& Vacca 1998). At about 5~Myrs WN stars clearly dominate
the WR star population and the strength of the \ion{C}{4} $\lambda$5808 line 
(indicating 
WC stars) decreases whereas the WR features at 
$\lambda$4660\AA ~(indicating WN stars) start to increase significantly.
The presences of both WN and WC stars therefore suggests an age
between 3 and 5 Myrs for the nuclear stellar population in NGC~6764
associated with the WR stars.

\subsection{\label{xxold}Modeling: General Remarks}

In our starburst model calculations, we used a Salpeter IMF with a lower mass
cut-off of 1 \solm ~and an upper mass cut-off of 120 \solm. The WR
stars signal that massive star formation has occured recently or is
still ongoing. On the other hand, an older stellar population that
produces most of the $K$ band continuum must be present. Therefore
only two simple scenorios can be considered producing the observed
properties: (1) Two starburst events 
describing discontinuous star formation due to episodic fueling
with molecular gas and (2) continuous star formation
assuming continuous fueling with molecular gas and/or a large nuclear 
reservoir of gas. 
\\
In a first step we tried to fit the diagnostic ratios for the two different
scenarios.  When the predicted \lk ~and \llyc ~luminosities
were matched to the observed values.
The spectral synthesis was used to address the non-stellar
properties of the nuclear region such as extinction, dust emission or 
AGN continuum contribution which also effect the overall values.
However, the derived values are rather small:
For the spectral synthesis fitting (which is independent of \lk, since it is
tied to the calibrated spectrum), we tried apparent extinction values of $A_V$=1$^{mag}$,
2$^{mag}$, 3$^{mag}$, 4$^{mag}$ and 5$^{mag}$. 
For the two star formation 
scenarios considered here, we find an extinction of (2 - 3)$^{mag}$ 
(mainly affecting the slope of the $K$ band spectrum at shorter wavelength end)
combined with a $\sim$ 5\% non-stellar contribution to the $K$ band continuum
which is either due to emission
from warm dust ($T$ $\sim$ 500 K) or AGN power law emission
(affecting the slope of the $K$ band spectrum at the longer
wavelength end).
We used the mean flux density at large radii (r $\sim$ 3.5'') to
estimate the contribution of the underlying old bulge population to
about 10 \% of the $K$ band continuum flux density in a 3''
aperture. However, the bulge population has a much larger spatial
extent than the stellar cluster mapped in its stellar absorption
lines.
\\
Both star formation possibilities are explored in the following.
The model parameters and results of the star forming events for the 
different scenarios are given in Tab. \ref{xxt04}.

\subsubsection{\label{xxsb}Two Young Starburst Events}

As discussed in section \ref{xxwr} the WR stars have an age of about 3
- 5 Myr. This low age immediately implies a second starburst or star
formation event that is responsible for the cool evolved stars indicated by 
the stellar absorption features (s. section \ref{xxco}). 
\\
\\
\noindent
{\it Starburst about 3 - 5~Myr ago (SB\#1):}
Assuming a starburst event with a decay time of 3~Myr 
seems reasonable given the fact that SN explosions might disturb the ISM
and prevent further star formation. 
For a starburst in that age range, STARS gives a mean ionizing 
hot star of type O7 to O5 with a
mean effective temperature of T$_{eff}$ $\approx$ 40000 - 45000K. 
Neglecting dilution effects (s. section
\ref{xxheh}) this corresponds to an HeI/Br$\gamma$ ratio of 0.7 (Lan\c{c}on
\& Rocca-Volmerage 1996). A comparison of $K$ band spectra from 
galaxies with AGN
and pure starburst galaxies (Vanzi et al. 1998) suggests that a strong HeI 
emission line is only present in the pure starburst galaxies without an AGN.
This probably indicates that most of the HeI line flux observed towards the
nucleus of NGC~6764 is due to young, hot stars and not due to an AGN
component. Under the assumption that all HeI line flux is associated
with the starburst, we can correct the observed HeI/Br$\gamma$ ratio of 
0.46$\pm$0.04 to obtain the Br$\gamma$ line flux associated with the 
starburst event. To fit the model ratio of 0.7 about 65\% of the observed
Br$\gamma$ line flux has to be due to the 5~Myr old starburst. This finding 
is in excellent agreement with the value
of 66\% for the H$\alpha$ line emission of Gon\c{c}alves, Veron-Cetty
\& Veron (1999). By fitting \llyc, we obtained the contibutrions of this
starburst to \lk, \lbol ~and \nsn.
The remaining differences in these quantities have to be attributed to the 
second starburst (or star formation) event and are used for the further
model fits.
\\
\\
\noindent
{\it Starburst about 25 Myr ago (SB\#2):}
The first starburst contributes only a few percent to the $K$ band
luminosity, and the old bulge population delivers about 10 \% (see
section \ref{xxold}).
Therefore, this second starburst has to account for about 85\% of the 
total \lk ~and the
total amount of \nsn ~without further large contributions to \llyc. 
A $\sim$ 15 Myr old starburst can account for most
of the remaining \llyc. However, if the age is about 25~Myr almost all of the
remaining 35\% of the Br$\gamma$ line flux could be associated with the 
LINER nucleus (s. Table \ref{xxt04}). 
At an age of 50~Myr, the
contribution of this starburst to the Lyman continuum is already
negligible and only small to the nuclear radio continuum.
We also tested
ages of $\sim$ 1~Gyr where AGB stars have a prominent contribution to
the $K$ band continuum (Lan\c{c}on 1999, Lan\c{c}on et al. 1999). This
possibility can be ruled out, since at that age the radio emission
could no longer be explained by star formation, i.e. synchrotron emission from 
SN remnants. 
The age for this second starburst event can not be so strict as
in the case of the WR starburst, since it is not possible to absolutely
constrain the contribution of this starburst to the $K$ band
luminosity, the Lyman continuum luminosity and the SN rate. However,
its age has to be larger than 15 Myr and below 50~Myr which is 
well below 1~Gyr.
Also, an underlying low-level constant star formation is
impossible, since such a population can not produce the \lk ~needed
without large contributions to \llyc ~(see section \ref{xxcs}).
\\
\\
\noindent
This scenario is quite opposite to the one found in IC~342 where a 70~pc
diameter starburst ring with an age of about 5~Myr surrounds a
nuclear starburst of about 15~Myr (B\"oker, F\"orster-Schreiber \& Genzel 
1997). In NGC~6764 the younger component ({\it SB\#1}) is concentrated
in the nucleus and surrounded or embedded in the older star formation
event ({\it SB\#2}). Higher angular resolution would allow us to
investigate if the {\it SB\#1} component is indeed concentrated on the
nucleus or is distributed in a small ring like in IC~342.
A comparison of the contribution of the two starburst events to the
optical continuum shows that the WR burst contributes about 15\% to
the synthesized stellar $V$ band continuum. This is in agreement with
the 11\% continuum contribution derived by Osterbrock \& Cohen (1982).

\subsubsection{\label{xxcs}Continuous Star Formation}

\begin{figure}[t!]
\begin{center}
\includegraphics[height=15.5cm,width=10.5cm,angle=-90.]{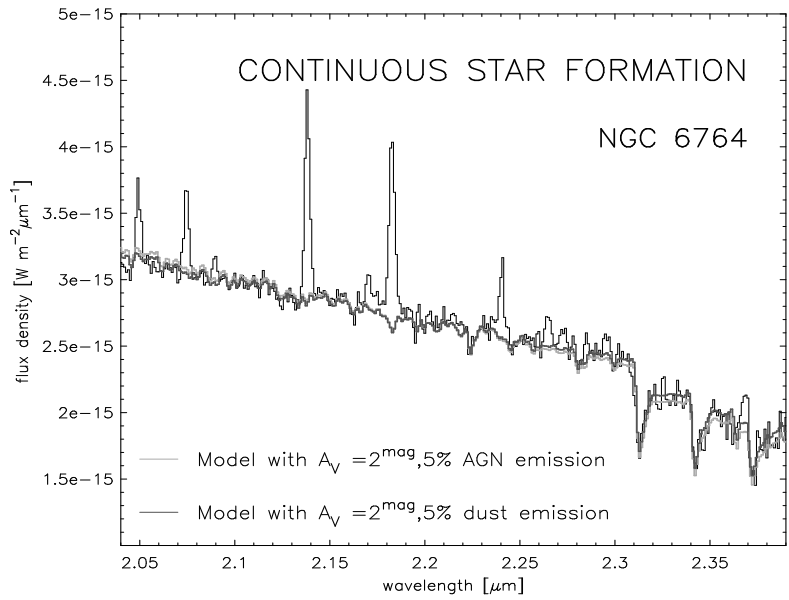}
\end{center}
\figcaption{\label{xxa07}
Comparison of the observed spectrum to the synthesized spectrum for
the case of the continuous star formation.
}
\end{figure}

If we assume continuous star formation in the nuclear region the nuclear stellar
population must have an age of $\geq$ 1 Gyr, since only then enough cool evolved
stars have been produced to obtain the observed ratio of L$_K$ to L$_{LyC}$.
It is necessary to reach at least the AGB phase for a reasonable
fit, since only then enough of the observed cool luminous stars are present. 
This analysis is, however, hampered by the fact, that the evolutionary
tracks of the high mass stars do not produce very cold red supergiants
(e.g. of type M) and that SPECSYN does not use spectra of AGB stars.
The spectra of AGB stars are approximated by those of a red supergiant
(RSG) of type M4Iab. This may affect the slope at the beginning of the
$K$ band, however, SPECSYN spectra start only at 2.0267 $\mu$m there
this effect is already weak. Since only a small
fraction of the total $K$ band light is due to AGB stars, this is
neglegible and for the purpose of the NIR spectral synthesis sufficient.
\\
The derived star formation rate (SFR) of this scenario is $\sim$ 0.3
\solm/yr. For an age of the stellar population of 1 Gyr this
translates into about 3$\times$10$^8$ \solm ~of molecular gas that
have been transformed into stars. 
The average star formation effiency (SFE) in giant molecular
cloud complexes (GMCs) is of the order of a few percent (Duerr et al. 1982,
Tenorio-Tagle \& Bodenheimer 1988), up to values of 30 - 40 \% only in the 
densest cores of these complexes (Lada 1982; see summary article by
Lada, Strom \& Myers 1993). If we assume an average value of up to
10\% for molecular clouds in starburst 
environments, the mass inflow rate needed is about 3 - 6 \solm/yr. 
Bars are
relative efficient in transporting gas down to small radii, Jogee, Kenney \&
Smith (1999) found a mass inflow rate of about 1 \solm/yr for the central
starburst in NGC~2782. However, Eckart et al. (1991) estimated a total
molecular gas mass of 2 - 6 $\times$10$^8$ \solm ~for NGC~6764.
Together with the fact that bars are short-lived features in the
evolution of galaxy ($\sim$ 10$^8$ yrs; Combes 1998), such a
high inflow rate over 1 Gyr seems highly unlikely. 
\\
Another aspect of 
this scenario is that the WR emission (HeI and Br$\gamma$ line) and the 
stellar absorption bands should have similar distributions even at highest 
spatial resolution. However, this is
already in contradiction to the observed spatial distribution of the H
and He recombination lines (unresolved) versus the extent of the
stellar absorption lines (similar to $K$ band continuum). 
It might be a possible scenario that due to dispersion of the nuclear stellar 
cluster the older stars have moved to larger radii.
A comparison to the upper limit of $\sim$ 3$\times$10$^9$ \solm ~for
the total mass in the inner 5'' (Eckart et al. 1991) shows that the
converted gas mass is only a small fraction ($\sim$ 10\%) and
therefore we expect the gravitational potential to minimize such an
effect. 
In addition, a continuous star formation would account for the
total H recombination line emission without any emission arising from
the AGN.
Therefore continuous star formation is higly unlikely, 
although the spectral fit (see Fig. \ref{xxa07}) is equally satisfying.

\section{SUMMARY AND CONCLUSIONS}

NGC6764 shows strong variations in its X-ray flux density by at least a 
factor of 2 on time-scales of 7 days.
This suggests the presence of a compact AGN with an upper size estimate
of $\rm R \approx \Delta t\ c \approx 1.6 \times 10^{16} cm$ or only about
10$^3$AU. In addition there is evidence for an
extended and possibly soft diffuse component of the X-ray emission.
The hot X-ray emitting gas and the synchrotron radiation
emitting relativistic gas component are spatially coexisting and are probably
both the result of strong recent nuclear star formation.
\\
NIR integral field spectroscopy of the nuclear region in the WR LINER
galaxy NGC~6764 reveals that the nuclear star formation is probably confined to
two areas, one of less than 100~pc (WR stars) and the other with an extent of
$\sim$ 200~pc (RSGs): 
The WR stars and the evolved cool stars are
not co-spatial and the younger starburst resides inside the older
starburst.
This picture is quite contrary to what is observed in many
starburst galaxies where the younger component is arranged in a ring
around the older nucleus (e.g. NGC~7552: Schinnerer et al. 1997,
IC~342: B\"oker, F\"orster-Schreiber \& Genzel 1997).
\\
Application of a population synthesis in conjunction with NIR spectral
synthesis infers an extinction towards the nuclear stellar cluster of
about (2 - 3)$^{mag}$ in agreement with the observed
Br$\delta$/Br$\gamma$ ratio and earlier findings by Eckart et al.
(1996). In addition we see evidence for a 5\% non-stellar contribution
to the $K$ band continuum either from warm dust or
a power law contribution from the AGN itself. 
In a 3'' aperture the $K$ band contribution of the bulge
population is about 10\%.
\\
The nuclear star formation history allows for two simple possibilities:
(1) Two starbursts with ages of 3 to 5~Myr and between 15 and $<$ 50~Myr and
decay times of 3~Myr which produce the WR stars and red supergiants that
contribute to a large amount of the $K$ band continuum.
(2) Continuous star formation with a SFR of $\sim$ 0.3 \solm/yr for at
least 1~Gyr.
\\
In the case of the 'two starburst' scenario,
an analysis of the H and He recombination lines shows that about 65\% of
the Br$\gamma$ line emission is associated with the young starburst,
leaving the rest for the LINER nucleus or/and the second starburst
event. Comparison of our data and data from the literature
with population synthesis results for WR dominated
cluster suggests an age of 3 - 5 Myr for the WR component in NGC~6764.
The age for the second starburst event is in the range of 15 Myr to
well below 1 Gyr depending on the contribution of this starburst to
the Lyman continuum and the radio continuum observed.
\\
The continuous star formation scenario seems highly unlikely given the
larger amount of molecular gas needed to be transported down to radii
of $\sim$ 160~pc over 1 Gyr. In addition, the spatial distribution 
of the line emission is expected to be cospatial with 
the stellar absorption lines - this is not observed.
Even if we allowed for an unresolved 35\% AGN contribution to the Br$\gamma$ 
such a case seems unlikely, because there is no indication 
for extended emission on a lower level in the radial profile of the Br$\gamma$ 
line emission.
Therefore, the most likely scenario is
that two recent starbursts have occured in the nuclear region of NGC~6764.
\\
The nucleus of NGC~6764 exhibits an interesting star formation
scenario, since the younger component seems
more compact than the older
one. One explanation could be that the bar is transporting the
molecular gas very close to the nucleus ($<$ 50~pc). In that case the
star formation might directly compete with the AGN for the fuel. It may also
influence or even control the AGN activity. The presence of a compact AGN 
(from the X-ray data) and violent recent nuclear star formation
underline the composite nature of the nucleus of NGC~6764.

\acknowledgments
We like to thank the MPE 3D group, especially N. Thatte, J.F.
Gallimore, M. Tecza and N.M. F\"orster-Schreiber, for their help taking 
the data. We also
thank the Calar Alto 3.5~m telescope staff for hospitality and support.
We are grateful to A. Sternberg for providing us with the population
synthesis code STARS. We also want to thank the anonymous referee for
her/his valuable comments which helped to improve the paper.

\clearpage

\begin{center}
{\Large \bf References}
\end{center}

\rf{Armus, L., Heckman, T.M., Miley, G.K., 1988, \apjjl{326}{L45}}
\rf{Baum, S.A., O'Dea, C.P., Dallacassa, D., de Bruyn, A.G., Pedlar,
    A., 1993, \apjj{419}{553}}
\rf{B\"oker, T., F\"orster-Schreiber, N.M., Genzel, R., 1997, \ajj {114}{1883}}
\rf{Boer, B., Schulz, H., 1989, in 'Extranuclear Activity in
    Galaxies', ESO Workshop, held in Garching, F.R.G., May 16-18, 1989,
    ed. E.J.A. Meurs and R.A.E. Fosbury; Publisher, European Southern
    Observatory, Garching bei M\"unchen, p. 299}
\rf{Condon, J. J., 1992, \arasa{30}{575}}
\rf{Boller, Th., 1999, Ap. \& S.S., 266, 49}
\rf{Conti, P.S., 1991, \apjj{377}{115}}
\rf{David, L.P., Harnden, F.R., Kearns, K.E., Zombeck, M.V., 1996, The
    ROSAT High Resolution Imager. SAO Press, Cambridge}
\rf{Dickey, J.M., Lockman, F.J., 1990, ARAA, 28, 215}
\rf{Doherty, R.M., Puxley, P.J., Lumsden, S.L., doyon, R., 1995, \mn {277}{577}}
\rf{Doyon, R., Joseph, R.D., Wright, G.S., 1994, \apjj{421}{101}}
\rf{Duerr, R., Imhoff, C.L., Lada, C.J., 1982, \apjj{261}{135}}
\rf{Eckart, A., Cameron, M., Boller, Th., Krabbe, A., Blietz, M.,
    Nakai, N., Wagner, S.J., Sternberg, A., 1996, \apjj{472}{588}}
\rf{Eckart, A., Cameron, M., Jackson, J.M., Genzel, R., Harris, A.I.,
    Wild, W., Zinnecker, H., 1990, \apjj {372}{67}}
\rf{Figer, D.F., McLean, I.S., Najarro, F., 1997, \apjj{486}{420}}
\rf{F\"orster-Schreiber, N.M., 2000, A.J., accepted}
\rf{Gehrz, R.D., Sramek, R.A., and Weedman, D.W., 1983, \apjj{267}{551}}
\rf{Genzel, R., Weitzel, L., Tacconi-Garman, Blietz, M., Krabbe, A.,
    Lutz, D., Sternberg, A., 1995, \apjj{444}{129}}
\rf{Gon\c{c}alves, A.C., Veron-Cetty, M.-P. \& Veron, P., 1999, \asas{135}{437}}
\rf{Jogee, S., Kenney, J.D.P., Smith, B.J., 1999, \apjj {526}{665}}
\rf{Kleinmann, S.G., Hall, D.N.B., 1986, \apjjs{62}{501}}
\rf{Krabbe, A., Sternberg, A., and Genzel, R., 1994, \apjj{425}{72}}
\rf{Kriss, G.A., Canizares, C.R., Ricker, G.R., 1980, \apjj {242}{492}}
\rf{Kunth, D., Contini, T., 1999, in 'Wolf-Rayet Phenomena in Massive
    Stars and Starburst Galaxies', Proc. IAU Symposium 193, ed. K.A. van
    der Hucht, G. Koenigsberger \& P.R.J. Eenens, p. 725}
\rf{Lada, E.A., 1992, \apjjl{393}{L28}}
\rf{Lada, E.A., Strom, K.M., Myers, P.C., 1993, in ''Protostars and
    Planets III'', p. 245}
\rf{Lan\c{c}on, A., 1999, 'Asymptotic Giant Branch Stars', IAU
    Symposium 191, in press, astro-ph9810474}
\rf{Lan\c{c}on, A., Mouhcine, M., Fioc, M., Silva, D., 1999, \asa {344}{L21}}
\rf{Lan\c{c}on, A., Rocca-Volmerange, B., 1996, New. A., 1, 215}
\rf{Larson, R. B., Tinsley, B. M., 1978, \apjj{219}{46}}
\rf{Leitherer, C. et al., 1999, \apjjs{123}{3}}
\rf{Leitherer, C., Heckman, T.M., 1995, \apjjs{96}{9}}
\rf{Lord, S., 1992, NASA Technical Memorandum 103957, Ames Research
    Center, Moffett Field, CA}
\rf{Maeder, A., Conti, P.S., 1994, ARAA, 32, 227}
\rf{Maiolino, R., Krabbe, A., Thatte, N., Genzel, R., 1998, \apjj {493}{650}}
\rf{Mas Hesse, J.M., and Kunth, D., 1991, Astr.Ap.Suppl. 88, 399}
\rf{Meynet, G., Maeder, A., Schaller, G., Schaerer, D.,
    Charbonnel, C., 1994, \asas{103}{97}}
\rf{Oliva, E., Origlia, L., 1998, \asa{332}{46}}
\rf{Oliva, E., Origlia, L., Kotilainen, J.K., Moorwood, A.F.M., 1995,
    \asa {301}{55}}
\rf{Origlia, L., Moorwood, A.F.M., Oliva, E., 1993, \asa {280}{536}}
\rf{Origlia, L., Oliva, E., 2000, \asa{357}{61}}
\rf{Osterbrock, D.E., 1989, Astrophysics of Gaseus Nebulae and
    Galactic Nuclei, University Science Books, Mill Valley, California}
\rf{Osterbrock, D.E., Cohen, R.D., 1982, \apjj{261}{64}}
\rf{Raymond, J.C., Smith, B.W., 1977, \apjs {35}{419}}
\rf{Rieke, G.H., 1978, \apjj {226}{550}}
\rf{Rieke, G. H., Lebofsky, M. J., Thompson, R. I., Low, F. J.,
   Tokunaga, A. T., 1980, \apjj{238}{24}}
\rf{Rieke, G. H., Loken, K., Rieke, M. J., Tamblyn, P., 1993, \apjj{412}{99}}
\rf{Rubin, VC., Thonnard, N., Ford, W.K., 1975, \apjj{199}{31}}
\rf{Sanders, D.B., Mirabel, I.F., 1996, \arasa{34}{749}}
\rf{Schaerer, D., Charbonnel, C., Meynet, G., Maeder, A. Schaller, G.,
   1993, \asas{102}{339}}
\rf{Schaerer, D., Contini, T., Pindao, M., 1999, \asas {136}{35}}
\rf{Schaerer, D., Vacca, W.D., 1998, \apjj{497}{618}}
\rf{Schinnerer, E., Eckart, A., Quirrenbach, A., B\"oker, T., Tacconi-Garman, 
    L.E., Krabbe, A., Sternberg, A., 1997, \apjj{488}{174}}
\rf{Schinnerer, E., Eckart, A., Tacconi, L.J., 1998, \apjj {500}{147}}
\rf{Schinnerer, E., Eckart, A., Tacconi, L.J. et al., 2000, in prep.}
\rf{Schmidt, M., Green, R.F., 1986, \apjj {305}{68}}
\rf{Telesco, C.M., Dressel, L.L., Wolstencroft, R.D., 1993, \apjj{414}{120}}
\rf{Tenorio-Tagle, G., Bodenheimer, P., 1988, \araa {26}{145}}
\rf{Thatte, N.A., Kroker, H., Weitzel, L., Tacconi-Garman, L.E.,
    Tecza, M., Krabbe, A., Genzel, R., 1995, Proc. SPIE Vol. 2475, 228}
\rf{Tr\"umper, J., 1983, Adv. Space Res., 4, 241}
\rf{Ulvestad, J.S., Wilson, A.S., Sramek, R.A., 1981, \apjj{247}{419}}
\rf{Vanzi, L., Alonso-Herrero, A., Rieke, G.H., 1998, \apjj {504}{93}}
\rf{Walter R., Fink H., 1993, \asa {274}{105}}
\rf{Weitzel, L., Krabbe, A., Kroker, H., Thatte, N., Tacconi-Garman,
    L. E., Cameron, M., Genzel, R., 1996, \asas {119}{531}}
\rf{Wilson, A.S., Willis, A.G., 1980, \apjj{240}{429}}

\clearpage

\begin{table}[htb]
\caption{\label{xxt01}Emission line fluxes in NGC~6764}
\begin{center}
\begin{tabular}{rrrrrr}\hline \hline
Aperture & Br$\delta$ & He I & Br$\gamma$ & Br$\delta$/Br$\gamma$ &
He I/Br$\gamma$\\
\[[''] & 1.945 $\mu$m & 2.058 $\mu$m & 2.166 $\mu$m &  & \\ \hline 
 2.00 &  3.37 $\pm$  0.48 &  2.14 $\pm$  0.16 &  4.64 $\pm$  0.20 &  0.73 $\pm$  0.11 &  0.46 $\pm$  0.04 \\ 
 3.00 &  3.90 $\pm$  0.72 &  2.93 $\pm$  0.24 &  6.40 $\pm$  0.29 &  0.61 $\pm$  0.12 &  0.46 $\pm$  0.04 \\ 
 4.00 &  4.95 $\pm$  0.94 &  3.00 $\pm$  0.31 &  7.59 $\pm$  0.38 &  0.65 $\pm$  0.13 &  0.39 $\pm$  0.05 \\ 
 5.00 &  5.71 $\pm$  1.21 &  3.30 $\pm$  0.40 &  8.20 $\pm$  0.49 &  0.70 $\pm$  0.15 &  0.40 $\pm$  0.05 \\ 
\hline \hline
\end{tabular}
\end{center}
Emission line fluxes are given in units of 10$^{-18}$ W m$^{-2}$ = 
10$^{-15}$ ergs cm$^{-2}$ s$^{-1}$. The given errors are the standard 
deviations (1$\sigma$). 
The lines were measured in the combined data
with an angular resolution of $\sim$ 1.2''.
\end{table}

\begin{table}[htb]
\caption{\label{xxt03}Absorption line equivalent widths in NGC~6764}
\begin{center}
\begin{tabular}{rrrrr}\hline \hline
Aperture & Na I & $^{12}$CO(2-0) & $^{12}$CO(3-1) & $^{13}$CO(2-0) \\
\[[''] &  2.206/2.209 $\mu$m & 2.294 $\mu$m & 2.323 $\mu$m & 2.345 $\mu$m \\ \hline
 2.00 &  2.81 $\pm$  0.71 & 13.29 $\pm$  1.46 & 14.52 $\pm$  2.58 &  8.61 $\pm$  3.53 \\ 
 3.00 &  2.86 $\pm$  0.68 & 13.09 $\pm$  1.40 & 13.72 $\pm$  2.48 & 10.06 $\pm$  3.39 \\ 
 4.00 &  2.70 $\pm$  0.70 & 13.70 $\pm$  1.47 & 14.61 $\pm$  2.60 & 11.04 $\pm$  3.55 \\ 
 5.00 &  1.66 $\pm$  0.75 & 14.06 $\pm$  1.61 & 13.64 $\pm$  2.85 & 11.54 $\pm$  3.90 \\ 
\hline \hline
\end{tabular}
\end{center}
Absorption line equivalent widths given in \AA. 
The given errors are the standard deviations (1$\sigma$).
The lines were measured in the combined data
with an angular resolution of $\sim$ 1.2''.
We used the definitions for the absorptions lines as given in Origlia
et al. 1993.
\end{table}

\begin{table}[htb]
\caption{\label{xxt04}Star Formation Scenarios in the Nucleus}
\begin{center}
\begin{tabular}{lrrrrrrr}\hline \hline
 & Obs. & SB \#1$^a$ & SB \#1$^b$ & SB \#2$^c$ & SB \#2$^d$ & SB \#2$^e$ &Cont. \\ \hline
L$_K$ [10$^7$ L$_{\odot}$] &       9.9 & 0.2 & 0.3 & 8.4 & 8.4 & 8.4 & 9.9 \\
L$_{LyC}$ [10$^8$ L$_{\odot}$]   & 3.9 & 2.6 & 2.6 & 1.4 & 0.2 & 0.0 & 4.1 \\
L$_{bol}$ [10$^9$ L$_{\odot}$]   & 9.0 & 1.3 & 2.1 & 6.3 & 7.5 & 5.4 & 6.3 \\
$\nu_{SN}$ [10$^{-2}$ yr$^{-1}$] & 1.2 & 0.0 & 0.1 & 1.4 & 3.0 & 0.1 & 0.5 \\
m$_{st}$ [10$^7$ M$_{\odot}$]    &     & 0.1 & 0.1 & 1.7 & 4.5 & 8.0 & 12.0 \\
m$_{gas}$ [10$^7$ M$_{\odot}$]   &     & 0.1 & 0.1 & 2.2 & 6.5 & 13.1 & 29.2 \\
\hline \hline
\end{tabular}
\end{center}
The observed values are derived assuming A$_V$=2$^{mag}$  
and using the following measured values in a
3'' aperture: S$_K$=6.74mJy, S$_{5GHz}$=3.4mJy,
F$_{Br\gamma}$=64.0$\times$10$^{-16}$ ergs cm$^{-2}$.
The equations can be found in Schinnerer, Eckart \& Tacconi
(1998). m$_{st}$ and m$_{gas}$ denote the present-day stellar mass and
the consumed gas mass to date, respectively.
\\
All starburst models are calculated using a Salpeter IMF with mass
cut-offs of 1 \solm ~to 120 \solm. 
\\
SB \#1$^a$: age of 3 Myr and decay time of 3 Myr (WR stars).
\\
SB \#1$^b$: age of 5 Myr and decay time of 3 Myr (WR stars).
\\
SB \#2$^c$: age of 15 Myr and decay time of 3 Myr (RSGs).
\\
SB \#2$^d$: age of 25 Myr and decay time of 3 Myr (RSGs).
\\
SB \#2$^e$: age of 50 Myr and decay time of 3 Myr (RSGs).
\\
Cont. : age $\geq$ 1 Gyr and continuous star formation. 
\end{table}

\clearpage

\clearpage

\clearpage

\clearpage

\clearpage

\clearpage

\end{document}